\documentclass[twocolumn,preprintnumbers,superscriptaddress,prl,nofootinbib]{revtex4-1}
\usepackage{graphicx}
\usepackage{times}
\usepackage{epsfig}
\usepackage{amsmath}
\usepackage{amsfonts}
\usepackage{amssymb}
\usepackage{url}
\usepackage{subfigure}
\usepackage{color}
\newcommand{\be}{\begin{equation}}
\newcommand{\ee}{\end{equation}}
\newcommand{\bea}{\begin{eqnarray}}
\newcommand{\eea}{\end{eqnarray}}

\begin{document}
\pagestyle{plain}
\title{Revisiting the direct detection of dark matter in simplified models}
\author{Tong Li}
\affiliation{
School of Physics, Nankai University, Tianjin 300071, China
}
\affiliation{
ARC Centre of Excellence for Particle Physics at the Terascale, School of Physics and Astronomy,
Monash University, Melbourne, Victoria 3800, Australia
}



\begin{abstract}
In this work we numerically re-examine the loop-induced WIMP-nucleon scattering cross section for the simplified dark matter models and the constraint set by the latest direct detection experiment. We consider a fermion, scalar or vector dark matter component from five simplified models with leptophobic spin-0 mediators coupled only to Standard Model quarks and dark matter particles. The tree-level WIMP-nucleon cross sections in these models are all momentum-suppressed. We calculate the non-suppressed spin-independent WIMP-nucleon cross sections from loop diagrams and investigate the constrained space of dark matter mass and mediator mass by Xenon1T. The constraints from indirect detection and collider search are also discussed.
\end{abstract}
\maketitle

The nature of dark matter (DM) remains unknown as one of the most mysterious components in our Universe. The search for dark matter is a crucial but challenging task for particle physics and astrophysics to affirm the properties of DM.
From particle physics point of view, the so-called weakly interacting massive particle (WIMP)~\cite{Steigman:1984ac} is believed to be one plausible cold dark matter candidate.
According to the different interacting methods between WIMP and ordinary matter, the WIMP search experiments are divided into three classes: direct detection (DD), indirect detection (ID) and collider search.

Among the three search methods, direct detection experiments search for the scattering of dark matter particles off atomic nuclei in underground detectors. Up to now, there has been no well-established observation of dark matter from direct detection experiments, leading instead to upper limits on the WIMP-nuclei scattering cross section.
The WIMP-nuclei elastic cross section can be divided into the spin-independent (SI) interaction and the spin-dependent (SD) interaction. The coherent sum of SI nucleon amplitudes leads to a strong enhancement for large nuclei, while there is no such enhancement for SD interactions in nuclei. Thus, it is simply argued that the restrictive constraint from DD experiments is stringent for SI but not for SD interacting cross section. However, Freytsis and Ligeti emphasized that one is very likely to underestimate the WIMP-nuclei scattering cross section when the leading-order contribution is SD or the leading-order SI contribution is highly momentum-suppressed~\cite{Freytsis:2010ne}.
In fact, the loop-level effect can induce distinct effective operators with non-suppressed SI amplitudes, after integrating out the loop diagrams~\cite{Freytsis:2010ne,Haisch:2013uaa,Ipek:2014gua,Arcadi:2017wqi,Sanderson:2018lmj,Herrero-Garcia:2018koq}.
Enhanced by the squared total nucleon number in a nucleus, the loop-level SI amplitudes can compensate the loop suppression and dominate the WIMP-nuclei cross section. Thus, recent null results from direct detection experiments should cast sizable constraint on such WIMP hypothesis with loop-induced SI cross section.

In this work, we numerically re-examine the loop-level SI cross section for the so-called simplified dark matter models and the constraint set by the latest direct detection experiments~\cite{Tan:2016zwf,Aprile:2017iyp}. The DD constraint is then compared with those from indirect detection and collider search. Specifically, we consider Dirac fermion, complex scalar and vector dark matter denoted by $\chi$, $\phi$ and $X$ respectively, with mediators only coupled to the Standard Model (SM) quarks and dark matter particles. This leptophobic framework was widely used to analyze dark matter searches in indirect detection, direct detection and collider experiments~\cite{Buchmueller:2013dya,Arina:2014yna,Alves:2015pea,Abdallah:2015ter,Boveia:2016mrp,Arina:2016cqj,Ismail:2016tod,Li:2016uph,Balazs:2017hxh,Li:2017nac,Morgante:2018tiq}.

In the simplified framework, the dark matter annihilation occurs through the exchange of either a spin-0 or spin-1 mediator in s-channel.
The minimal flavor violation indicates that the couplings of vector and axial-vector quark bilinears are chosen to be universal and those of scalar and pseudo-scalar bilinears, i.e. $\bar{q}q$ and $\bar{q}\gamma_5 q$, are scaled by SM quark mass $m_q$~\cite{Buras:2000dm,Goodman:2010ku}. The spin-1 mediator scenario via vector or axial-vector interaction is thus highly confined by the $Z'\to$ dijet search at the Large Hadron Collider (LHC)~\cite{Khachatryan:2016ecr,Sirunyan:2016iap,Sirunyan:2017nvi,Sirunyan:2018wcm}. We thus consider the spin-0 mediator scenario
with $m_q^2$ suppression at the LHC, coming from the SM-like Yukawa coupling. Among the structure combinations in the spin-0 mediator scenario, the tree-level WIMP-nuclei scattering cross sections of only five forms are either momentum-suppressed SI or SD, designated by D2, D3, D4, C and V using the notation of effective field theories (EFTs)~\cite{Goodman:2010ku}, as shown in Table~\ref{tab:operator}. Moreover, for these models, the annihilation cross sections are mostly not velocity-suppressed, rendering relatively strong ID constraint. We thus investigate the loop-induced direct detection constraint on simplified dark matter models D2, D3, D4, C and V, and compared with the ID constraint and LHC limit.

\begin{table}[h]
\begin{center}
\begin{tabular}{|c|c|c|c|}
        \hline
        Interations & ID & DD & Collider\\
        \hline
        ${\rm D2}: \bar{\chi}\gamma_5\chi\oplus\bar{q}q$ & $\sigma v\sim \mathcal{O}(1)$ & $\sigma_{\rm SI}\sim \mathcal{O}(q^2)$ & $\mathcal{O}(m_q^2)$ \\
        \hline
        ${\rm D3}: \bar{\chi}\chi\oplus\bar{q}\gamma_5q$ & $\sigma v\sim \mathcal{O}(v^2)$ & $\sigma_{\rm SD}\sim \mathcal{O}(q^2)$ & $\mathcal{O}(m_q^2)$ \\
        \hline
        ${\rm D4}: \bar{\chi}\gamma_5\chi\oplus\bar{q}\gamma_5 q$ & $\sigma v\sim \mathcal{O}(1)$ & $\sigma_{\rm SD}\sim \mathcal{O}(q^4)$ & $\mathcal{O}(m_q^2)$\\
        \hline
        ${\rm C}: \phi^\dagger\phi\oplus\bar{q}\gamma_5 q$ & $\sigma v\sim \mathcal{O}(1)$ & $\sigma_{\rm SD}\sim \mathcal{O}(q^2)$ & $\mathcal{O}(m_q^2)$ \\
        \hline
        ${\rm V}: X^\dagger_\mu X^\mu\oplus\bar{q}\gamma_5 q$ & $\sigma v\sim \mathcal{O}(1)$ & $\sigma_{\rm SD}\sim \mathcal{O}(q^2)$ & $\mathcal{O}(m_q^2)$\\
        \hline
\end{tabular}
\end{center}
\caption{Interactions considered in this work and their tree-level scaling effects for ID, DD and collider search.
}
\label{tab:operator}
\end{table}


We first describe the simplified dark matter models in Table~\ref{tab:operator}.
The dark matter particles ($\chi,\phi,X$) couple to the SM quarks through a spin-0 mediator $S_2$, $S_3$, $S_4$, $S_C$ or $S_V$, corresponding to structure D2, D3, D4, C or V respectively. The corresponding interactions are as follows~\cite{Berlin:2014tja,Abdallah:2015ter}
\begin{eqnarray}
&&{\cal L}_{\rm D2} = -ig_{\chi}^{\rm D2} S_2\bar{\chi}\gamma_5 \chi - S_2\sum_{q}g_q^{\rm D2} {m_q\over v_0}\bar{q} q,
\label{eq:interactionD2}\\
&&{\cal L}_{\rm D3} = -g_{\chi}^{\rm D3} S_3\bar{\chi} \chi - iS_3\sum_{q}g_q^{\rm D3} {m_q\over v_0}\bar{q}\gamma_5 q,
\label{eq:interactionD3}\\
&&{\cal L}_{\rm D4} = -ig_{\chi}^{\rm D4} S_4\bar{\chi}\gamma_5 \chi - iS_4\sum_{q}g_q^{\rm D4} {m_q\over v_0}\bar{q}\gamma_5 q,
\label{eq:interactionD4}\\
&&{\cal L}_{\rm C} = -g_{\phi}^{\rm C} m_\phi S_C \phi^\dagger\phi - iS_C\sum_{q}g_q^{\rm C} {m_q\over v_0}\bar{q}\gamma_5 q,
\label{eq:interactionC}\\
&&{\cal L}_{\rm V} = -g_{X}^{\rm V} m_X S_V X^\dagger_\mu X^\mu - iS_V\sum_{q}g_q^{\rm V} {m_q\over v_0}\bar{q}\gamma_5 q,
\label{eq:interactionV}
\end{eqnarray}
where $v_0=246$ GeV is the Higgs vacuum expectation value. Following the general choice in the analysis of dark matter searches in literatures, we take $g_{\chi}^{\rm D2}=g_{\chi}^{\rm D3}=g_{\chi}^{\rm D4}=g_q^{\rm D2}=g_q^{\rm D3}=g_q^{\rm D4}=g_{\phi}^{\rm C}=g_{X}^{\rm V}=g_q^{\rm C}=g_q^{\rm V}=1$ in the calculations below. Under the above assumption, these dark matter models are described by two parameters, i.e. the dark matter mass $m_{\rm DM}=m_\chi, m_\phi, m_X$ and the mediator mass $m_{\rm Med}=m_{S_2}$, $m_{S_3}$, $m_{S_4}$, $m_{S_C}$ or $m_{S_V}$.

\begin{figure}[h!]
\begin{center}
\includegraphics[scale=1,width=6cm]{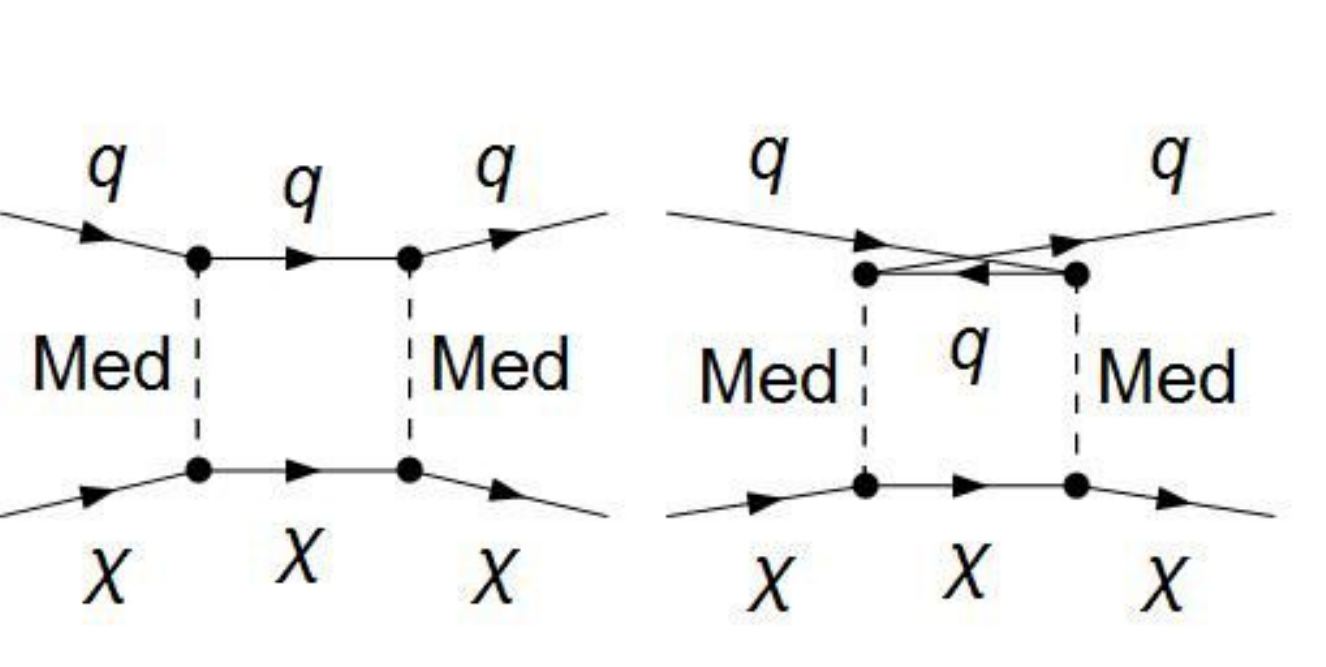}
\end{center}
\caption{Box loop diagrams generating the non-suppressed SI WIMP-nucleon cross section. The mediator is generally designated by ``Med''. The Dirac fermionic DM $\chi$ here can also be replaced by complex scalar $\phi$ or vector $X$.
}
\label{figLoop}
\end{figure}

For the above simplified models, WIMP-quark scattering amplitudes, through the box loop diagrams as shown in Fig.~\ref{figLoop}, essentially generate the generic dimension-6 operators in Table~\ref{tab:dmqint} which give non-suppressed SI WIMP-nucleon cross sections.
Because of the spin-0 mediator scenario we consider, the obtained operators are scaled by SM-like Yukawa couplings as follows, for scalar interactions for example
\begin{eqnarray}
\sum_q {m_q^2\over v_0^2} \mathcal{C}(m_{\rm DM},m_{\rm Med},m_q) \ \overline{\rm DM} \ {\rm{DM}} \ \bar{q}q,
\label{DMqint}
\end{eqnarray}
where $\mathcal{C}$ is the coefficient function after integrating out the loop diagrams. $\rm{DM}$ and $\overline{\rm DM}$ here represent the DM particles $\chi, \phi, X$ and their anti-particles, respectively. As a result, the vector interactions leading to only valence quarks contributions for WIMP-nucleon amplitudes and the valence quarks contributions to scalar interactions are both negligible. We thus only consider the above scalar interactions from heavy quarks contribution, through the relation between the heavy quark component of the nucleus and the gluon condensate~\cite{Drees:1993bu}.

\begin{table}[h]
\begin{center}
\begin{tabular}{|c|c|c|}
        \hline
        DM & scalar int. & vector int.\\
        \hline
        fermion $\chi$ & $\bar{\chi}\chi\bar{q}q$ & $\bar{\chi}\gamma_\mu\chi \bar{q}\gamma^\mu q$ \\
        \hline
        scalar $\phi$ & $2m_\phi \phi^\dagger \phi \bar{q}q$ & $\phi^\dagger \overleftrightarrow{\partial_\mu}\phi \bar{q}\gamma^\mu q$ \\
        \hline
        vector $X_\mu$ & $2m_X X_\mu^\dagger X^\mu \bar{q}q$ & $X_\nu^\dagger \overleftrightarrow{\partial_\mu}X^\nu \bar{q}\gamma^\mu q$\\
        \hline
\end{tabular}
\end{center}
\caption{The WIMP-quark interactions generating non-suppressed SI WIMP-nucleon cross section, for fermion, scalar and vector DM.
}
\label{tab:dmqint}
\end{table}

Converting WIMP-quark amplitudes in Eq.~(\ref{DMqint}) to WIMP-nucleon amplitudes, in terms of the nucleon form factor $f_{\rm TG}\approx 0.894$ for the gluon condensate~\cite{Drees:1993bu,MO}, we obtain the loop-induced WIMP-nucleon SI cross section for the simplified models we consider
\begin{eqnarray}
&&\sigma_{{\rm DM} N}^{\rm SI}={\mu^2_{\rm{DM} N}\over \pi}|\mathcal{I}_i(m_{\rm DM},m_{\rm Med})|^2,\\
&&\mathcal{I}_i(m_{\rm DM},m_{\rm Med})\approx \sum_{q=c,b,t}{m_q m_N\over v_0^2}{2\over 27}f_{\rm TG} \mathcal{C}_{i}(m_{\rm DM},m_{\rm Med},m_q),\nonumber \\
\end{eqnarray}
where $i=$ D2, D3, D4, C, V and $\mu_{\rm{DM} N}=m_{\rm DM} m_N/(m_{\rm DM}+m_N)$ is the reduced mass with $m_N$ being the nucleon mass. The detailed $\mathcal{C}_i$ coefficients in expression of Passarino-Veltman functions~\cite{Passarino:1978jh} are collected in Appendix.

\begin{figure}[h!]
\begin{center}
\includegraphics[scale=1,width=4.1cm]{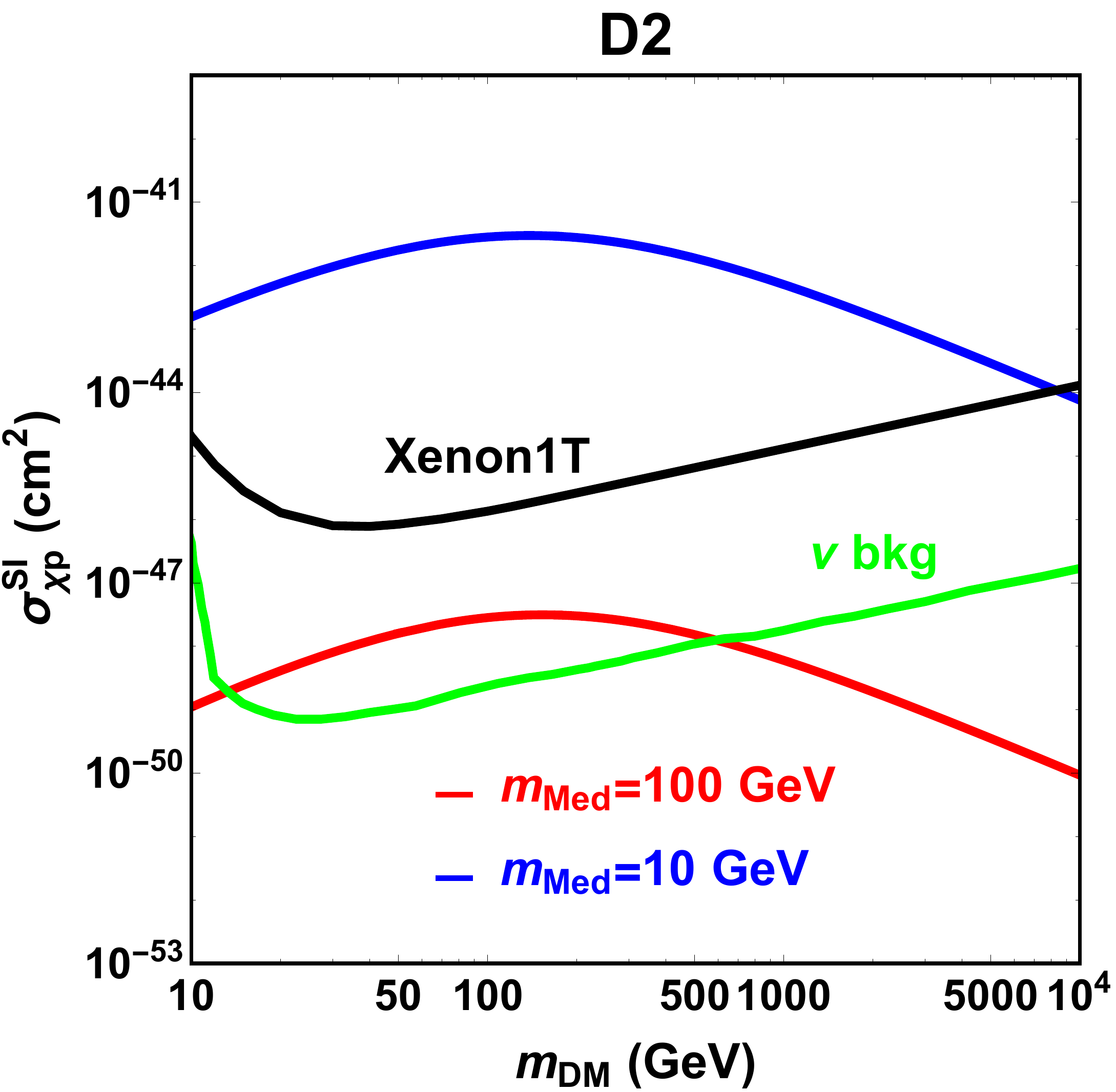}
\includegraphics[scale=1,width=4cm]{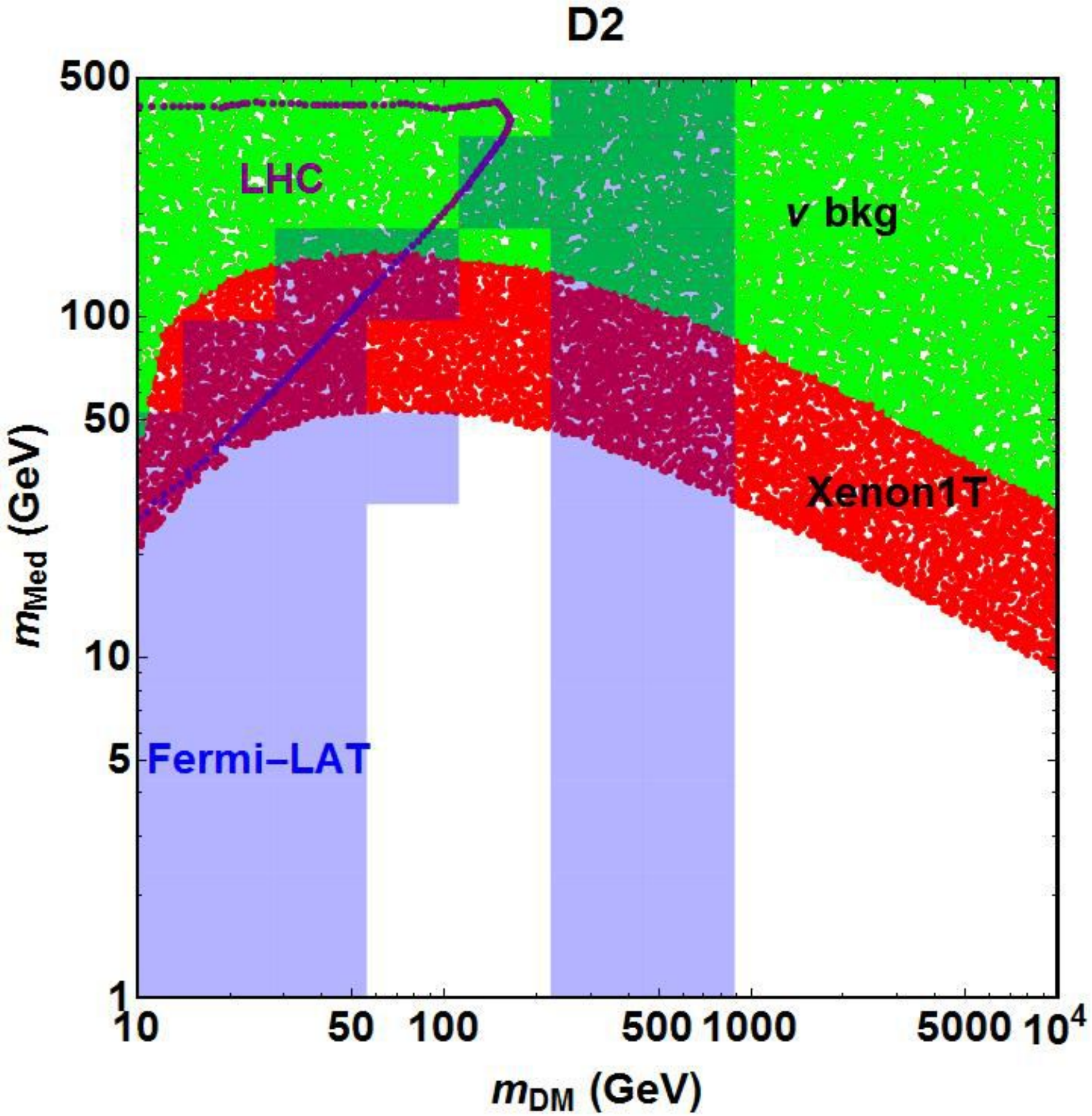}
\end{center}
\caption{Left: Spin-independent DM-proton scattering cross section at one loop level versus DM mass for D2 interaction. We assume the mediator mass to be 10 GeV or 100 GeV. The Xenon1T limit~\cite{Aprile:2017iyp} and the neutrino background are also shown. Right: The allowed region of $m_{\rm Med}$ vs. $m_{\rm DM}$ by Xenon1T (red) and the region below the neutrino floor (green). The region on the left-hand side of the purple curve is excluded by the LHC. The blue values are excluded by the measurement of dwarf galaxies by Fermi-LAT.
}
\label{figD2}
\end{figure}

\begin{figure}[h!]
\begin{center}
\includegraphics[scale=1,width=4.1cm]{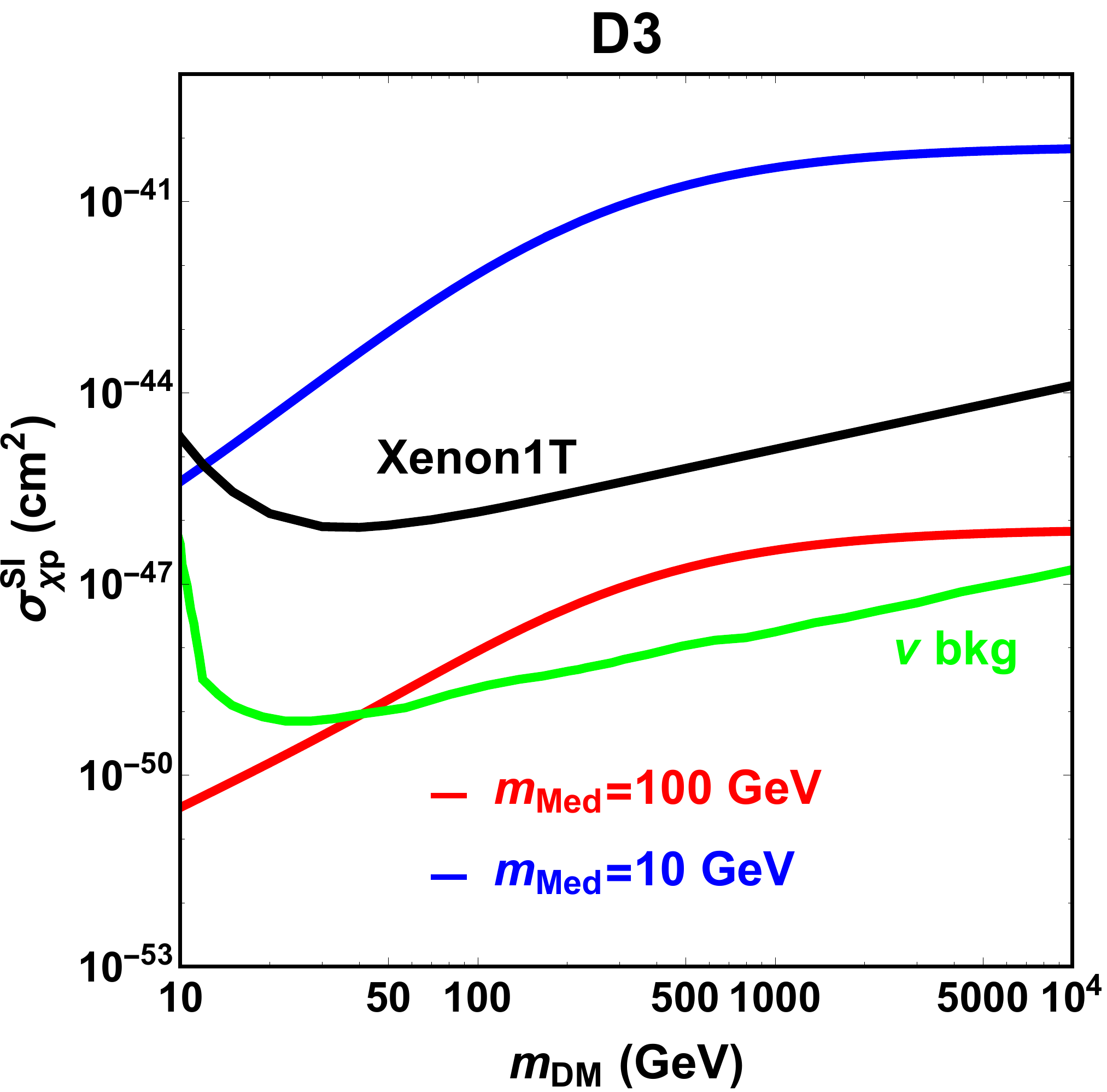}
\includegraphics[scale=1,width=4cm]{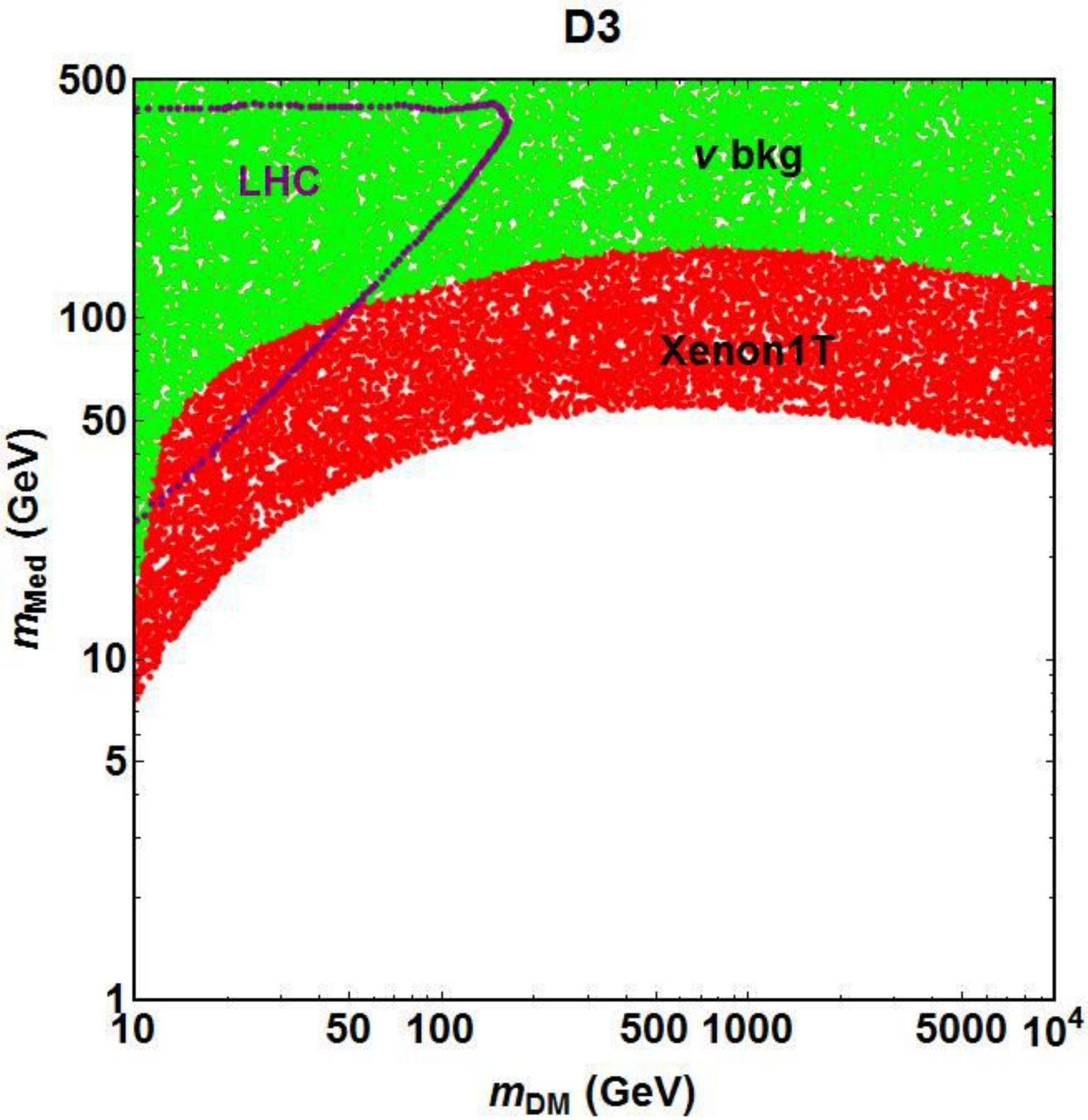}
\end{center}
\caption{Results for D3 interaction as labeled in Fig.~\ref{figD2}.
The region excluded by Fermi-LAT is not shown because of the low-velocity suppressed annihilation cross section for D3~\cite{Berlin:2014tja,Balazs:2017hxh}.
}
\label{figD3}
\end{figure}

\begin{figure}[h!]
\begin{center}
\includegraphics[scale=1,width=4.1cm]{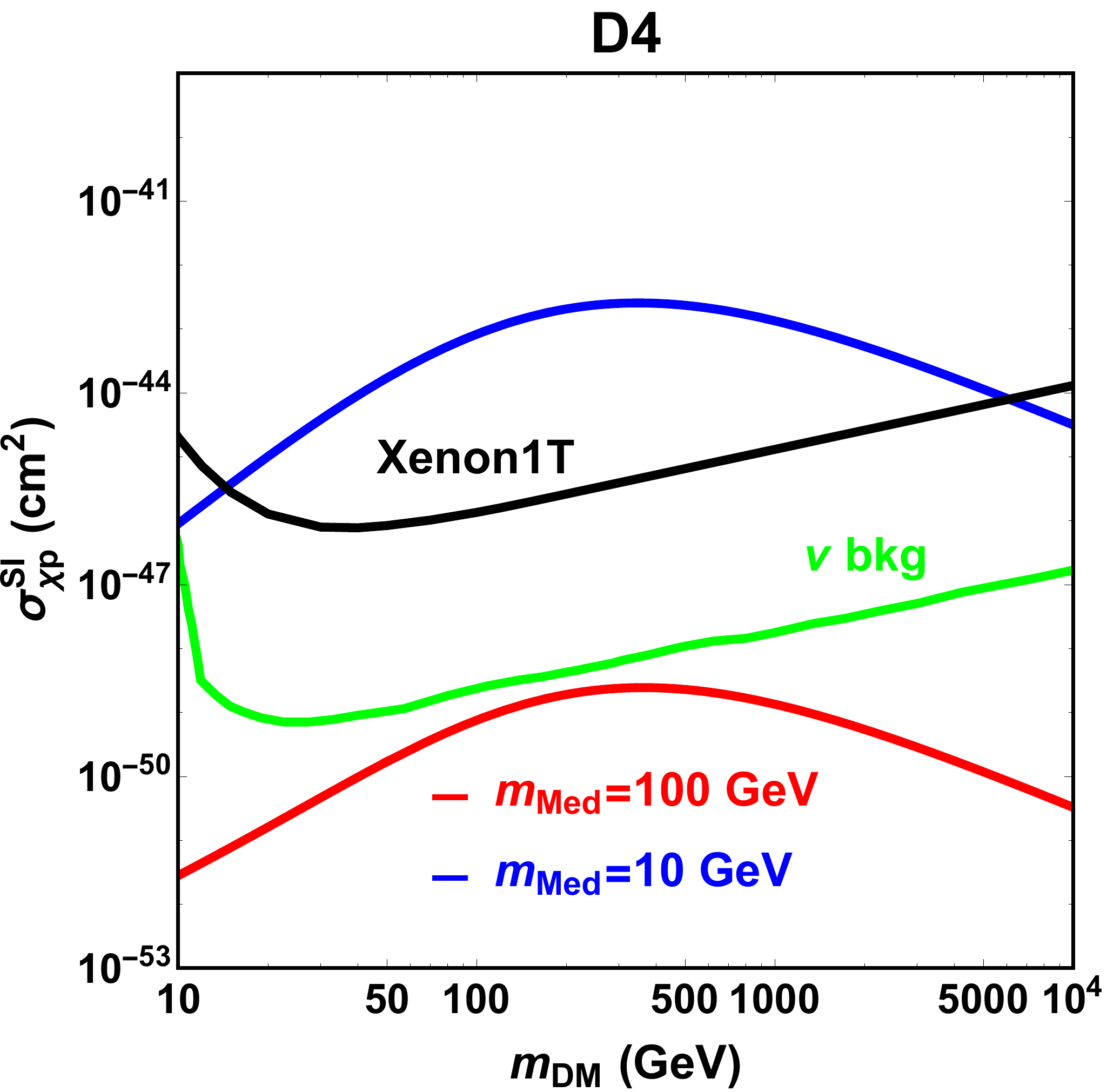}
\includegraphics[scale=1,width=4cm]{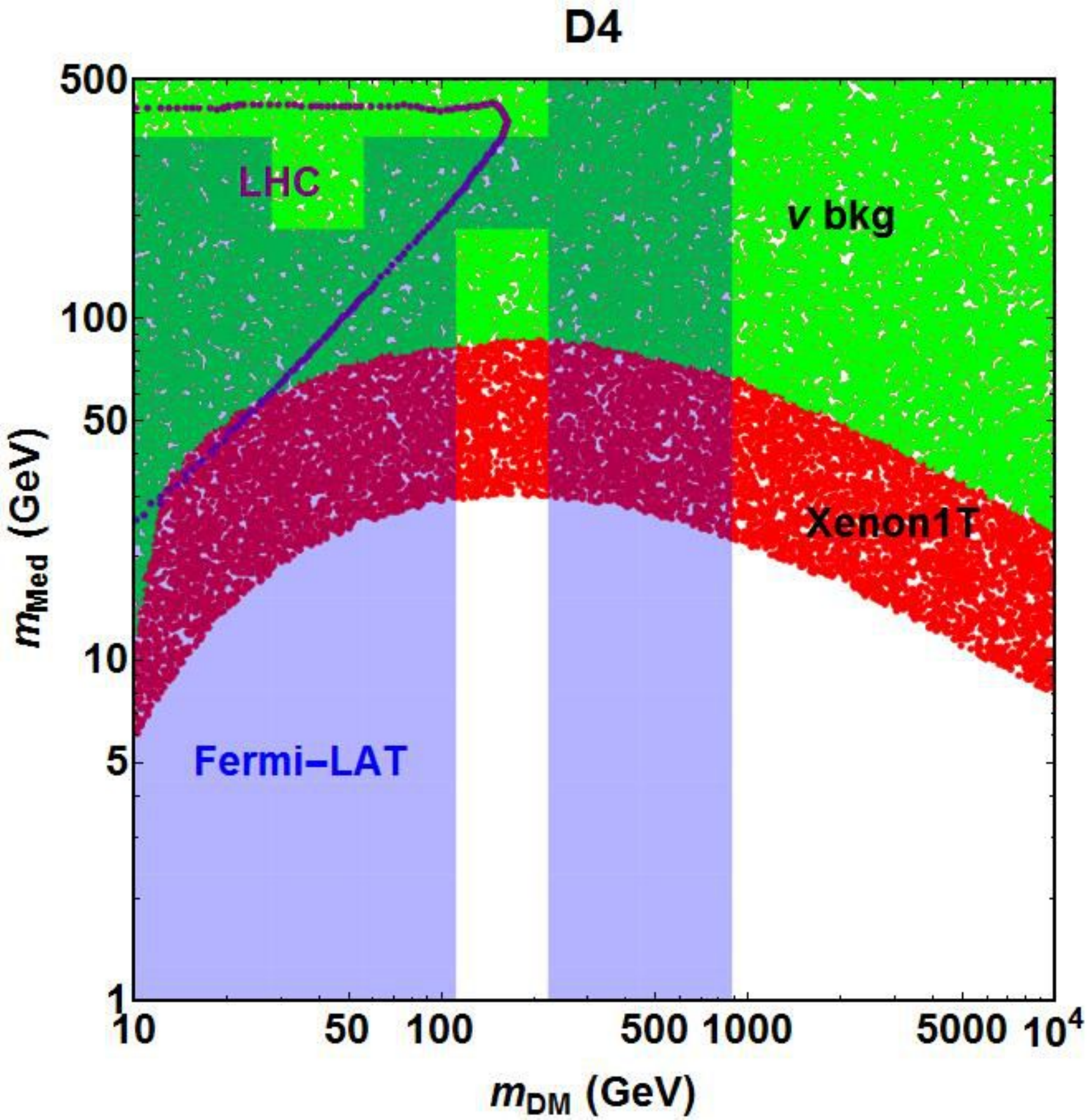}
\end{center}
\caption{Results for D4 interaction as labeled in Fig.~\ref{figD2}.
}
\label{figD4}
\end{figure}

\begin{figure}[h!]
\begin{center}
\includegraphics[scale=1,width=4.1cm]{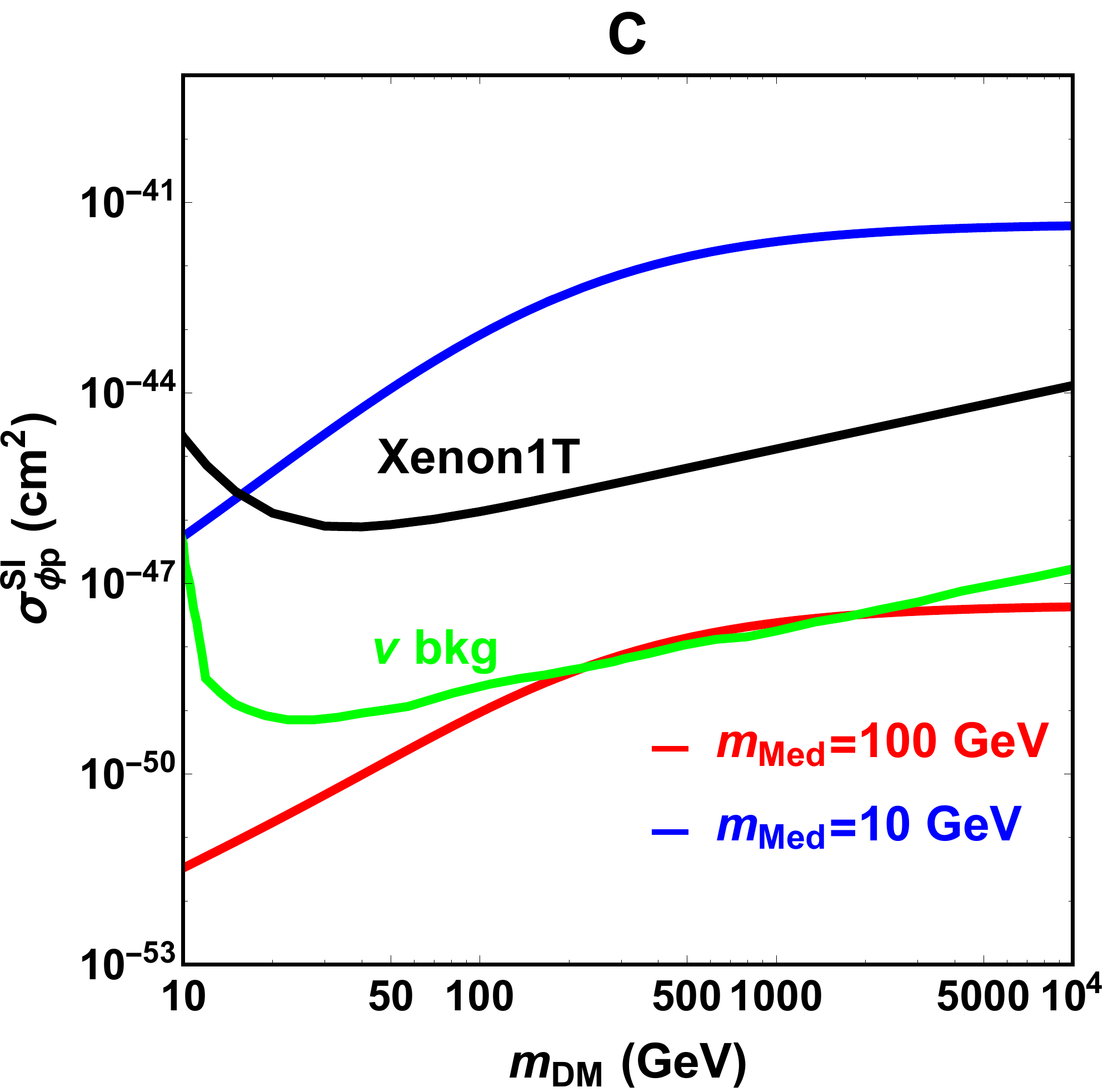}
\includegraphics[scale=1,width=4cm]{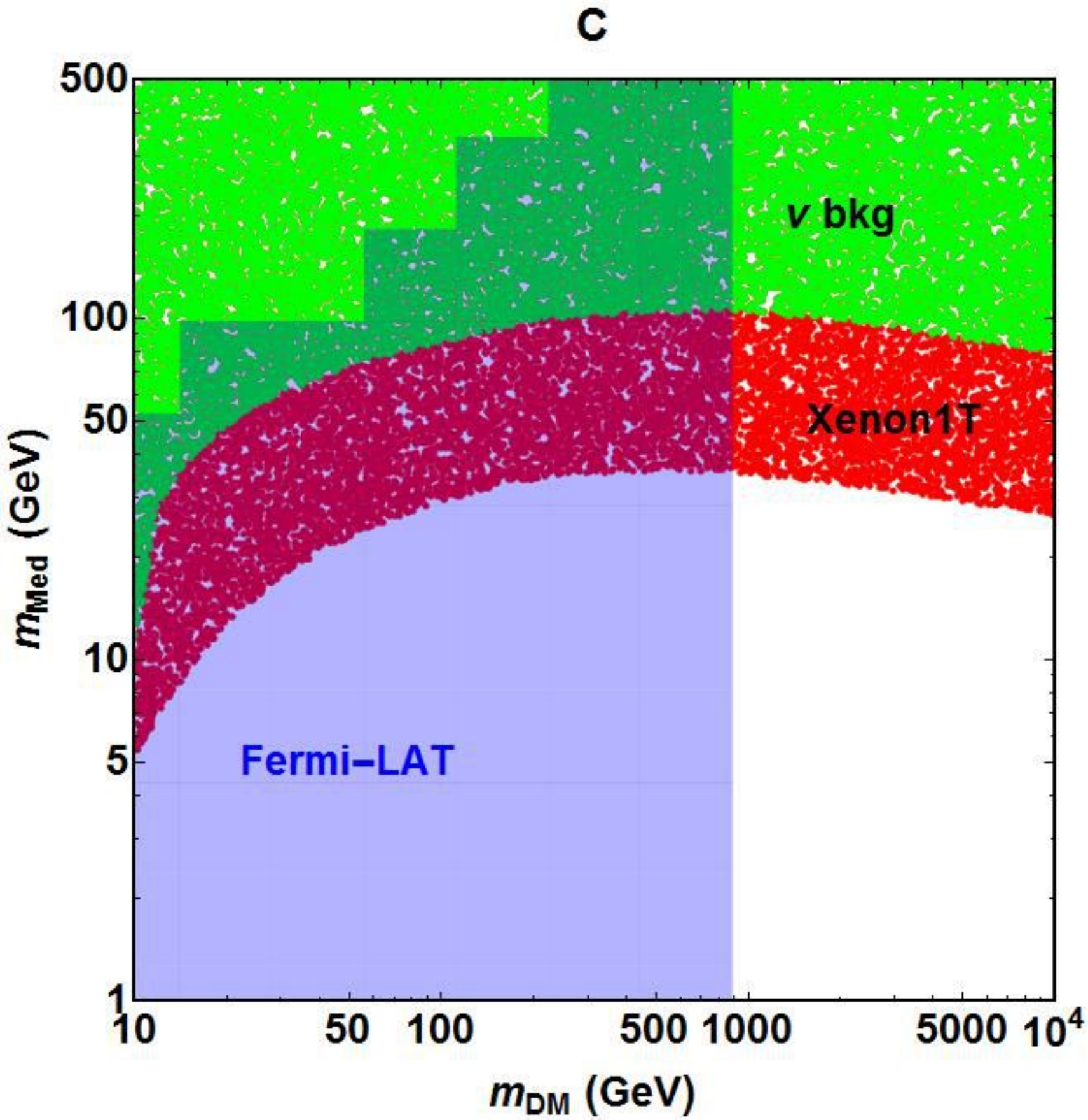}
\end{center}
\caption{Results for C interaction as labeled in Fig.~\ref{figD2}.
}
\label{figC}
\end{figure}

\begin{figure}[h!]
\begin{center}
\includegraphics[scale=1,width=4.1cm]{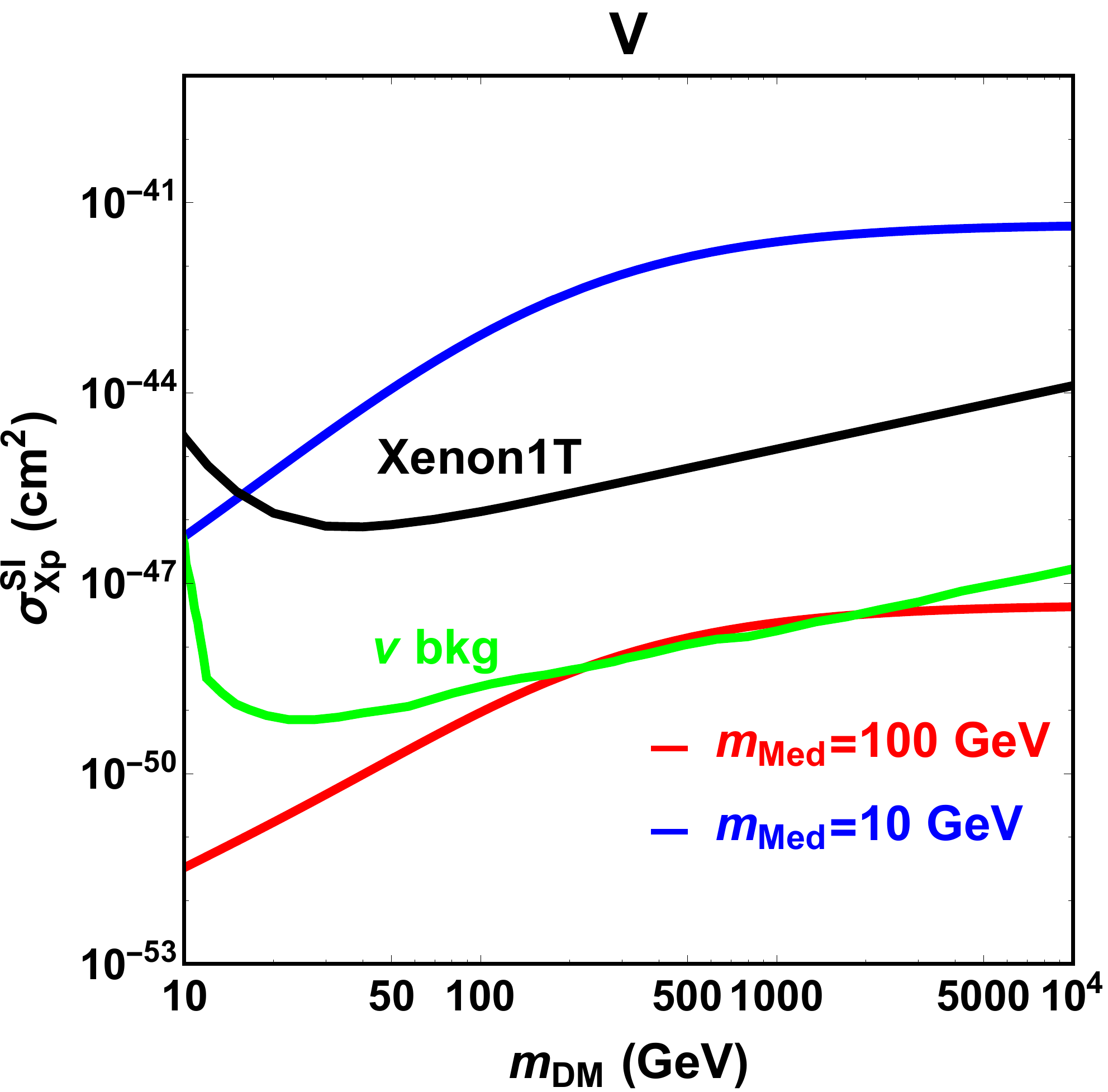}
\includegraphics[scale=1,width=4cm]{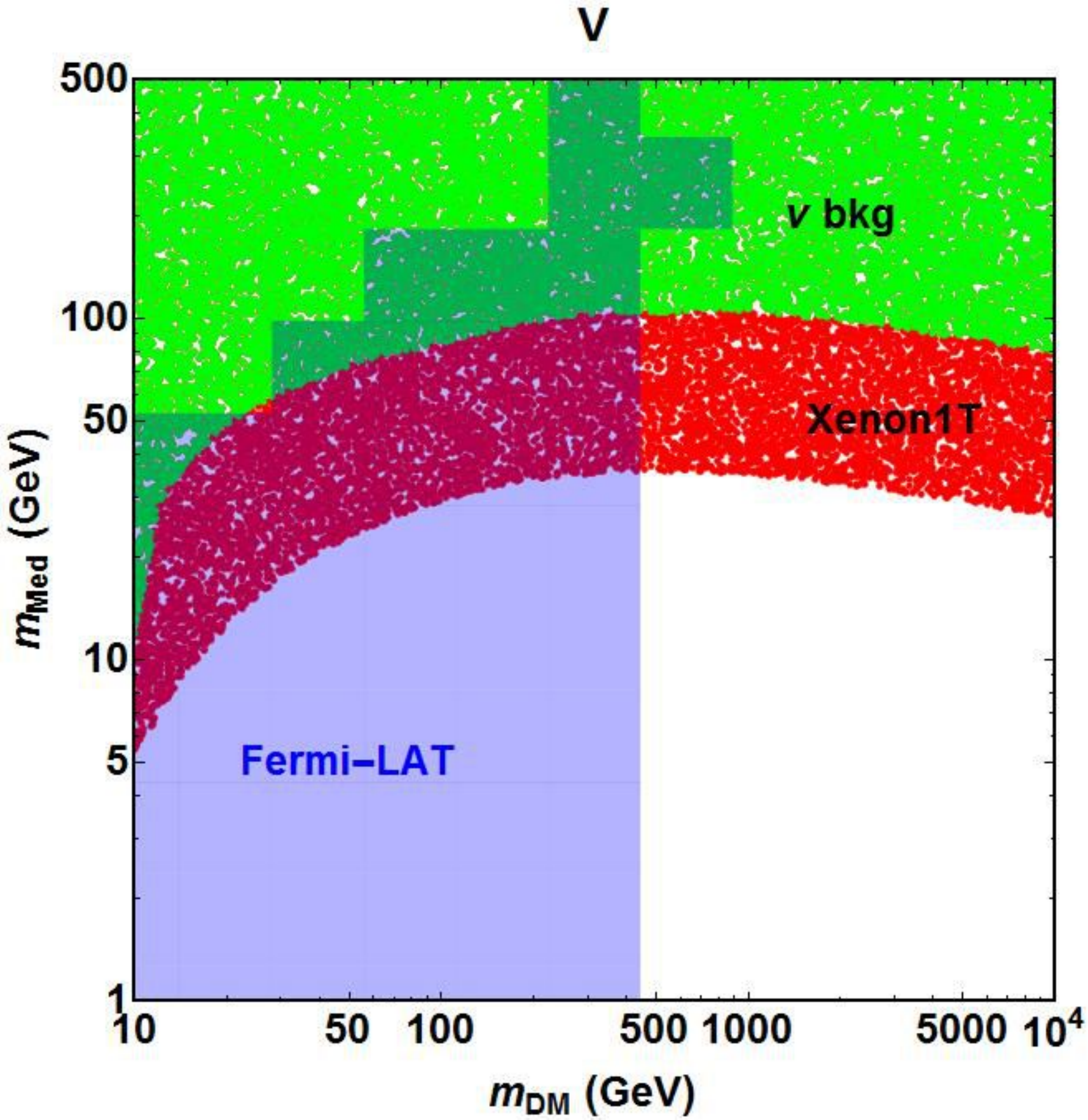}
\end{center}
\caption{Results for V interaction as labeled in Fig.~\ref{figD2}.
}
\label{figV}
\end{figure}

For each model, we produce the FeynRules~\cite{Alloul:2013bka} files followed by FeynArts~\cite{Hahn:2000kx} to generate the loop diagrams. The diagrams are then analytically implemented by FeynCalc~\cite{Mertig:1990an,Shtabovenko:2016sxi} and
we use Package-X~\cite{Patel:2015tea} to calculate the obtained Passarino-Veltman functions. In Figs.~\ref{figD2}, \ref{figD3}, \ref{figD4}, \ref{figC} and \ref{figV}, we display the results of SI DM-proton scattering cross section generated at loop-level for models D2, D3, D4, C and V, respectively. The left panels are for the scattering cross sections as a function of DM mass with different mediator masses. As expected, heavier mediator can generally reduce the WIMP-nucleon cross section. For fixed mediator mass, the Xenon1T limit can be evaded for models D3, C and V with relatively light DM or for models D2 and D4 with both low and high DM masses. These mass regions are yet more likely to lead to cross sections below the neutrino floor. The allowed regions of $m_{\rm Med}$ versus $m_{\rm DM}$ by Xenon1T (red) and the regions with $\sigma^{\rm SI}$ below the neutrino floor (green) are shown in the right panels. We find that mediator masses as high as about 30-50 GeV are excluded by Xenon1T for $m_{\rm DM}\simeq 100$ GeV in models D2 and D4 or for $m_{\rm DM}\gtrsim 100$ GeV in models D3, C and V.
For the low DM mass region in all models and the high DM mass region only in models D2 and D4, mediator masses around 5-10 GeV still survive.

In these figures, we also show the constraints from indirect detection and collider search on the mass space.
Dwarf galaxies with a large amount of dark matter are the bright targets to search for gamma rays from dark matter annihilation.
The Fermi Large Area Telescope (LAT) has searched for gamma ray emission from the dwarf spheroidal satellite galaxies (dSphs) of the Milky Way and detected no excess. Fermi-LAT thus placed upper limit on the dark matter annihilation cross section from a combined analysis of multiple Milky Way dSphs~\cite{Ackermann:2015zua,Fermi-LAT:2016uux}. For individual dwarf galaxy target, Fermi-LAT tabulated the delta-log-likelihood values as a function of the energy flux bin-by-bin.
The gamma ray energy flux from dark matter annihilation for $j$th energy bin is given by
\begin{eqnarray}
\Phi^E_{j,k}(m_{\rm DM},\langle \sigma v\rangle,J_k)=\frac{\langle \sigma v\rangle}{16\pi m_{\rm DM}^2}J_k\int^{E^{\rm max}_j}_{E^{\rm min}_j}E\frac{dN_\gamma}{dE}dE,
\end{eqnarray}
where $J_k$ is the J factor for $k$th dwarf. The energy flux is only dependent on $m_{\rm DM}$, $\langle \sigma v\rangle$ and $J_k$, and calculable for any dark matter annihilating process induced by the above simplified models. The likelihood for $k$th dwarf is
\begin{eqnarray}
&&\mathcal{L}_k(m_{\rm DM},\langle \sigma v\rangle,J_k)\nonumber \\
&&=\mathcal{L}_J(J_k|\bar{J}_k,\sigma_k)\prod_j \mathcal{L}_{j,k}(\Phi^E_{j,k}(m_{\rm DM},\langle \sigma v\rangle,J_k)),
\end{eqnarray}
where $\mathcal{L}_{j,k}$ is the tabulated likelihood provided by Fermi-LAT for each dwarf and calculated energy flux and the uncertainty of the J factors is taken into account by profiling over $J_k$ in the likelihood below
\begin{eqnarray}
\mathcal{L}_J(J_k|\bar{J}_k,\sigma_k)&=&{1\over \ln(10)J_k\sqrt{2\pi}\sigma_k}\nonumber \\
&\times& e^{-(\log_{10}(J_k)-\log_{10}(\bar{J}_k))^2/2\sigma_k^2},
\end{eqnarray}
with the measured J factor $\bar{J}_k$ and error $\sigma_k$. This likelihood form was adopted by Fermi-LAT collaboration in Eq.~(3) of Ref.~\cite{Ackermann:2015zua}. A joint likelihood for all dwarfs is then performed as
\begin{eqnarray}
\mathcal{L}(m_{\rm DM},\langle \sigma v\rangle,\mathbb{J})=\prod_k \mathcal{L}_k(m_{\rm DM},\langle \sigma v\rangle,J_k),
\end{eqnarray}
where $\mathbb{J}$ is the set of J factors $J_k$. In our implementation we adopt the corresponding values of $\mathcal{L}_{j,k}$ and $\bar{J}_k, \sigma_k$ for 19 dwarf galaxies considered in Ref.~\cite{Fermi-LAT:2016uux}.

As Fermi-LAT found no gamma ray excess from the dSphs, one can set upper limit on the annihilation cross section for a given $m_{\rm DM}$ by taking J factors as nuisance parameters in the maximum likelihood analysis. Following Fermi's approach, the delta-log-likelihood is given by
\begin{eqnarray}
-2\Delta \ln \mathcal{L}(m_{\rm DM},\langle \sigma v\rangle)=-2\ln\left({\mathcal{L}(m_{\rm DM},\langle \sigma v\rangle,\widehat{\widehat{\mathbb{J}}})\over \mathcal{L}(m_{\rm DM},\widehat{\langle \sigma v\rangle},\widehat{\mathbb{J}})}\right),
\end{eqnarray}
where $\widehat{\langle \sigma v\rangle}$ and $\widehat{\mathbb{J}}$ maximize the likelihood at any given $m_{\rm DM}$, and $\widehat{\widehat{\mathbb{J}}}$ maximize the likelihood for given $m_{\rm DM}$ and $\langle \sigma v\rangle$.
The 95\% C.L. upper limit on annihilation cross section for a given $m_{\rm DM}$ is determined by demanding $-2\Delta \ln\mathcal{L}(m_{\rm DM},\langle \sigma v\rangle)\leq 2.71$. We perform the likelihood analysis using Minuit~\cite{James:1975dr}.
Once the annihilation cross section calculated by a certain set of $m_{\rm DM}$ and $m_{\rm Med}$ is larger than the limit, we claim the corresponding mass values in Figs.~\ref{figD2}, \ref{figD3}, \ref{figD4}, \ref{figC} and \ref{figV} are excluded by Fermi-LAT dSphs (blue values). The Fermi-LAT dSphs exclude a majority of mass parameters for $m_{\rm DM}\lesssim 1$ TeV and $m_{\rm Med}\lesssim 500$ GeV, in models D2, D4, C and V. For models D2 and D4, in particular, they have survived region around $m_{\rm DM}\simeq 100$ GeV due to not kinematically allowed $\chi\bar{\chi}\to {\rm Med}\to t\bar{t}$ channel and low-velocity suppressed annihilation to mediator pairs~\cite{Balazs:2017hxh,Li:2017nac}.

On the other hand, LHC performed DM search using events with large missing energy plus energetic jet for pseudo-scalar mediator model (D4) at $\sqrt{s}=13$ TeV collisions~\cite{Sirunyan:2017jix}. The exclusion limit is presented in the plane of mediator mass versus dark matter mass. As models D2, D3 and D4 are closely related operators, they should have very similar collider constraints and thus we adopt the LHC constraint on model D4 for D2 and D3 as well (denoted by purple curves). One can see that LHC only excludes a small triangle region with $m_{\rm DM}\lesssim 200$ GeV and $30 \ {\rm GeV}\lesssim m_{\rm Med}\lesssim 400$ GeV for these models.
Besides the dedicated exclusion limit from LHC collaborations, there exist multiple studies of LHC constraints on the simplified DM models from theory groups, as seen in Refs.~\cite{Arina:2016cqj,Banerjee:2017wxi} and the references therein. They allow the mediator coupling strengths to the SM quarks and DM pairs to vary in a broader range and give stringent exclusion limit for the couplings larger than 1.
In addition to the limit from mono-jet search, when the mediator couples to the third generation SM fermions, strong constraint can be imposed on the simplified models from the search for new resonance decaying into heavy SM fermions or photons. According to Ref.~\cite{Banerjee:2017wxi}, from $t\bar{t}$ resonance search, one obtains the upper limit on the pseudoscalar mediator mass as about 400 GeV.
As there has been no dedicated exclusion limit from LHC collaboration, we omit the bounds for scalar and vector DM candidates here. One should keep in mind that the exclusion limits for them should be similar to the fermionic DM models displayed above, as the relevant element for the LHC searches is the nature of the mediator but not the DM.

One should keep in mind that the operator coefficients should in principle be evaluated at the scale
where direct detection scattering occurs, by performing Renormalization Group (RG) evolution from the high energy theory defined at the mediator mass scale~\cite{DEramo:2014nmf}. The operator mixing due to the RGE flow can also generate a contribution leading to spin-independent scattering cross section.
According to Ref.~\cite{Crivellin:2014qxa}, the evolution matrix for the scalar interaction leads to a non-vanishing mixing between operator $\bar{\chi}\chi\bar{q}q$ and $\bar{\chi}\chi G_{\mu\nu}G^{\mu\nu}$ with $G_{\mu\nu}$ being the gluon field strength tensor. This mixing gives a contribution to the coefficient of $\bar{\chi}\chi\bar{q}q$ which yields a suppression of about $1/(12\pi)$ compared to the loop-induced coefficient considered above. Thus, this RGE contribution has a numerically negligible impact on $\sigma^{\rm SI}_{\chi N}$. Interesting RGE contribution only takes place for the direct detection with vector mediator~\cite{DEramo:2014nmf,DEramo:2016gos}.

In summary, motivated by the SI WIMP-nucleon cross section from the loop-level effect and the latest direct detection experiments, we revisit the direct detection of simplified dark matter models. We consider fermion, scalar or vector dark matter candidate from five simplified models with leptophobic
spin-0 mediators coupled only to SM quarks and dark matter particles via scalar and/or pseudo-scalar bilinear. The tree-level WIMP-nucleon cross sections in these models are all momentum-suppressed. We calculate the non-suppressed SI WIMP-nucleon cross sections from loop diagrams and investigate the allowed region of DM mass and mediator mass by Xenon1T. We find that, including the loop-level SI cross sections, the sensitivity of direct detection to these models is complementary to the constraints from indirect detection and collider search.

\section*{Acknowledgments}
This work is supported in part by the ARC Centre of Excellence for Particle Physics at the Terascale.

\section*{Appendix}
For $q(p_1) {\rm DM}(p_2)\to q(k_1) {\rm DM}(k_2)$ process with Mandelstam variables as $s=(p_1+p_2)^2$, $t=(p_1-k_1)^2$ and $u=(p_1-k_2)^2$,
the Passarino-Veltman functions for ${\rm D2}, {\rm D3}, {\rm D4}, {\rm C}, {\rm V}$ are
\begin{eqnarray}
&&\mathcal{C}_{\rm D2}={m_q m_\chi\over 16\pi^2}\nonumber \\
&&[2D_2(m_q^2,s,m_\chi^2,t,m_\chi^2,m_q^2;m_{\rm Med}^2,m_q^2,m_\chi^2,m_{\rm Med}^2)+\nonumber \\
&&2D_2(m_q^2,u,m_\chi^2,t,m_\chi^2,m_q^2;m_{\rm Med}^2,m_q^2,m_\chi^2,m_{\rm Med}^2)+\nonumber \\
&&D_{12}(m_q^2,s,m_\chi^2,t,m_\chi^2,m_q^2;m_{\rm Med}^2,m_q^2,m_\chi^2,m_{\rm Med}^2)+\nonumber \\
&&D_{12}(m_q^2,u,m_\chi^2,t,m_\chi^2,m_q^2;m_{\rm Med}^2,m_q^2,m_\chi^2,m_{\rm Med}^2)],
\end{eqnarray}

\begin{eqnarray}
&&\mathcal{C}_{\rm D3}={m_q m_\chi\over 16\pi^2}\nonumber \\
&&[2D_1(m_q^2,s,m_\chi^2,t,m_\chi^2,m_q^2;m_{\rm Med}^2,m_q^2,m_\chi^2,m_{\rm Med}^2)+\nonumber \\
&&2D_1(m_q^2,u,m_\chi^2,t,m_\chi^2,m_q^2;m_{\rm Med}^2,m_q^2,m_\chi^2,m_{\rm Med}^2)+\nonumber \\
&&D_{12}(m_q^2,s,m_\chi^2,t,m_\chi^2,m_q^2;m_{\rm Med}^2,m_q^2,m_\chi^2,m_{\rm Med}^2)+\nonumber \\
&&D_{12}(m_q^2,u,m_\chi^2,t,m_\chi^2,m_q^2;m_{\rm Med}^2,m_q^2,m_\chi^2,m_{\rm Med}^2)],
\end{eqnarray}

\begin{eqnarray}
&&\mathcal{C}_{\rm D4}=-{m_q m_\chi\over 16\pi^2}\nonumber \\
&&[D_{12}(m_q^2,s,m_\chi^2,t,m_\chi^2,m_q^2;m_{\rm Med}^2,m_q^2,m_\chi^2,m_{\rm Med}^2)+\nonumber \\
&&D_{12}(m_q^2,u,m_\chi^2,t,m_\chi^2,m_q^2;m_{\rm Med}^2,m_q^2,m_\chi^2,m_{\rm Med}^2)],
\end{eqnarray}
(We check that $\mathcal{C}_{\rm D4}$ agrees with the result of pseudoscalar mediator model in Ref.~\cite{Arcadi:2017wqi})

\begin{eqnarray}
&&\mathcal{C}_{\rm C}=-{m_q m_\phi\over 32\pi^2}\nonumber \\
&&[D_{1}(m_q^2,s,m_\phi^2,t,m_\phi^2,m_q^2;m_{\rm Med}^2,m_q^2,m_\phi^2,m_{\rm Med}^2)+\nonumber \\
&&D_{1}(m_q^2,u,m_\phi^2,t,m_\phi^2,m_q^2;m_{\rm Med}^2,m_q^2,m_\phi^2,m_{\rm Med}^2)],
\end{eqnarray}

\begin{eqnarray}
&&\mathcal{C}_{\rm V}={m_q m_X\over 32\pi^2}\nonumber \\
&&[D_{1}(m_q^2,s,m_X^2,t,m_X^2,m_q^2;m_{\rm Med}^2,m_q^2,m_X^2,m_{\rm Med}^2)+\nonumber \\
&&D_{1}(m_q^2,u,m_X^2,t,m_X^2,m_q^2;m_{\rm Med}^2,m_q^2,m_X^2,m_{\rm Med}^2)].
\end{eqnarray}


\end{document}